\documentclass[twocolumn,amsmath,amssymb]{revtex4}
\usepackage[T1]{fontenc}
\usepackage{amsmath,graphicx,amssymb,epsfig,babel,dsfont,amscd,ulem,tikz}
\usepackage{mathrsfs}
\usepackage{color}
\usepackage{placeins}
\usepackage{hyperref}
\usepackage{physics}

\renewcommand{\Re}{{\rm Re}}
\renewcommand{\Im}{{\rm Im}}
\newcommand{\rd}{{\rm d}}

\newcommand{\ri}{{\rm i}}

\newcommand{\re}{{\rm e}}

\newcommand{\rIP}{\rm IP}
\newcommand{\rOP}{\rm OP}
\newcommand{\rlong}{\rm long}
\newcommand{\rtrans}{\rm trans}

\newcommand{\alphamat}{\underline{\underline{\alpha}}}

\begin{document}

\title{Substrate-driven topological engineering in plasmonic Su-Schrieffer-Heeger chains}

\author{F. Herz}
\affiliation{Institut f\"{u}r Physik, Carl von Ossietzky Universit\"{a}t, 26111, Oldenburg, Germany}

\author{A. Naeimi}
\affiliation{Institut f\"{u}r Physik, Carl von Ossietzky Universit\"{a}t, 26111, Oldenburg, Germany}

\author{S.-A. Biehs}
\email{s.age.biehs@uol.de}
\affiliation{Institut f\"{u}r Physik, Carl von Ossietzky Universit\"{a}t, 26111, Oldenburg, Germany}

\date{\today}

\begin{abstract}
We demonstrate the possibility of engineering the topological band structure of a plasmonic 
Su-Schrieffer-Heeger (SSH) chain through the interaction with its electromagnetic environment. 
We find that the long-range interaction of the in-plane modes of the SSH chain with the surface 
plasmon polaritons of a planar substrate introduces a band hybridization
connected to a change of the Zak phase. On the other hand, the short-range interaction with 
the substrate introduces a band touching, again with a change in the Zak phase. Surprisingly, 
this second mechanism enables the emergence of topologically protected edge modes for parameters 
which correspond to the topologically trivial phase for an isolated plasmonic SSH chain. 
We study these mechanisms by changing the chain-substrate distance and the dimerization parameter.
Finally, we discuss the robustness against disorder and, as one example, the impact of the observed effects
on the near-field radiative heat transfer and the local density of states along the chain. 
Our findings pave the way to the engineering of edge modes in plasmonic topological 
configurations via the coupling to a plasmonic environment.
\end{abstract}

\maketitle

%
%

\section{Introduction}\label{sec:intro}

The field of topological photonics, pioneered by the seminal theoretical work of Haldane and Raghu \cite{Raghu2008, Haldane2008} and later realized experimentally by Wang et al. \cite{Wang2009}, has opened new avenues for robust light manipulation. While many studies have focused on complex 2D lattices \cite{Xie2018, Ozawa2019}, ranging from magneto-optical photonic crystals \cite{Skirlo2015} and coupled resonator arrays \cite{Hafezi2013} to waveguide lattices \cite{Rechtsman2013} and pure dielectric metamaterials \cite{Khanikaev2013, Wu2015}, the Su-Schrieffer-Heeger (SSH) model \cite{Su1979} remains a cornerstone for understanding 1D topological insulators. 

Recently, these concepts have been extended to nanoparticle chains, where the discrete nature of the constituents allows for tailored light-matter interactions. In zigzag-shaped plasmonic chains, it was demonstrated that topological edge states emerge, which obey the same mathematical framework as Majorana fermions in superconducting quantum wires~\cite{Poddubny2014}, with topological phases depending strictly on the polarization. Similarly, by drawing analogies to topological materials such as Weyl semimetals, topological invariants have been identified in bipartite nanoparticle chains. These systems host topologically protected edge modes according to the bulk-boundary correspondence~\cite{Ling2015, Downing2017}, which exhibit remarkable resilience against structural disorder and retardation effects~\cite{Downing2018, Pocock2018}. This inherent robustness, combined with the capacity for long-range interactions, makes SSH chains particularly attractive for controlling radiative heat transfer as first highlighted in Ref.~\cite{Ott2020_2} (see also reviews in Refs.~\cite{Liu2024, Didari-Bader2024}).

Consequently, it has been shown that edge modes can serve as the dominant heat transfer channel~\cite{Ott2020_2, Wang2023_2}, persisting in quasi-periodic configurations~\cite{Wang2023}, effectively modulate the heat flux between specific particles within a chain~\cite{Gong2024}, and influence the internal temperature evolution to generate steep gradients~\cite{Nikbakht2023}. In such systems, the local density of states is enhanced~\cite{Ott2021}, a phenomenon that is predicted to be detectable with near-field thermal imaging microscopes~\cite{Herz2022}. In 2D systems, it has been shown theoretically that there can be an energy transport predominantly through surface-like edge states rather than bulk modes, allowing for controlling the direction of the radiative heat flow~\cite{OTTHoney}. For practical experimental implementations, the role of the environment is crucial. Recent studies suggest that coupling edge modes to the surface plasmon polaritons (SPPs) of a substrate can facilitate long-range radiative heat transfer~\cite{Naeimi2026}, while the use of non-reciprocal materials allows for directional control of the heat flux~\cite{Yang2025}. However, precise active control over these topological advantages remains a challenge. Early strategies have focused on the doping-dependence of resonance frequencies in silicon-based chains~\cite{Wang2020} or the influence of photonic cavities~\cite{Downing2019}.
 
In this work, we investigate how the proximity of an SSH chain to a substrate surface can be exploited to engineer the topological band structure and induce a topological phase transition. We explicitly show how changes in the dimerization and the chain-substrate distance induce two distinct phase-transition mechanisms and how they affect the evolution of the band topology. Furthermore, we analyze the robustness of the edge modes with respect to positional disorder. These effects are not only important for experimental setups where chains are placed on substrates but also provide a versatile method to tune edge mode properties by modifying the environment rather than the chain geometry itself. As an application, we consider the radiative heat flux along an SSH chain. In Ref.~\cite{Naeimi2026}, we demonstrated that the radiative heat flux can be long-range and significantly enhanced for dimerization parameters $\beta > 1$ due to the coupling between edge modes and surface modes. Here, we rather focus on the heat flux to illustrate the impact of the two distinct phase-transition mechanisms. In particular, the emergence of edge modes for $\beta < 1$ is completely novel, as is our investigation into the robustness of the radiative heat flux against positional disorder. This setup is also promising for applications involving scattering-type microscopes \cite{Efrima1979, Wiesendanger1994, Keilmann2004}, near-field thermal imaging microscopes \cite{Herz2022}, thermal management in nanostructures \cite{Guha2012}, enhanced resonance energy transfer \cite{Velizhanin2012, Biehs2013, Poudel2016}, local density of states (LDOS) engineering \cite{Messina2013B, Ott2021}, spontaneous emission \cite{Cortes2012}, and entanglement control \cite{Bellomo2015}, among other specific examples.

The paper is structured as follows: After presenting the theoretical framework in Ch.~\ref{sec:Framework}, we analyze the band structure and edge modes of the coupled SSH-chain-substrate system in Ch.~\ref{sec:band}. Therein, we examine the dependence of the band structure on the dimerization parameter in Ch.~\ref{sec:beta} and on the gap size in Ch.~\ref{sec:z}, alongside the substrate doping concentration. Additionally, the robustness of the topologically protected edge modes against structural disorder is discussed in Ch.~\ref{sec:robust}, while we discuss the impact on the radiative heat transfer and the LDOS as examples in Ch.~\ref{sec:RHT}. Finally, we summarize our findings in Ch.~\ref{sec:conclusion}.

%
%
\section{Theoretical framework}\label{sec:Framework}

\begin{figure}
	\centering
	\includegraphics[]{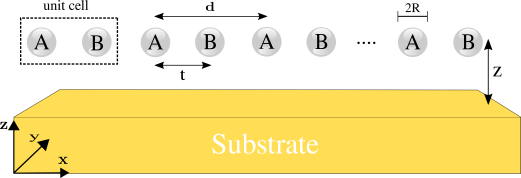}
	\caption{Sketch of the considered system. A bipartite SSH chain of identical nanoparticles of radius $R$ with lattice constant $d$ in each sublattice $A$ and $B$ and separation distance $t$ within each unit cell is placed at a distance $z$ in close vicinity to a semi-infinite planar substrate.}
	\label{Fig:Sketch}
\end{figure}

The system under study, illustrated in Fig.~\ref{Fig:Sketch}, comprises a bipartite array of $N$ identical uniform spherical nanoparticles with radius $R$ and polarizability $\alpha$ positioned at a distance $z$ from a planar surface. We distinguish between sublattices A and B with a common lattice period $d$. For the numerical calculations, we choose $R = 100$ nm. The spatial arrangement is governed by the intra-cell separation $t$ and the inter-cell spacing $d - t$, which we link through the dimerization parameter $\beta$ such that $t = \beta d / 2$. While $\beta = 1$ represents a perfectly periodic chain with uniform spacing, variations in $\beta$ allow us to tune the relative interaction strengths: for $\beta < 1$, the dipole-dipole interaction is primarily enhanced within the unit cell, whereas for $\beta > 1$, the coupling between particles of adjacent unit cells becomes dominant. Note that for an infinite chain, dimerizations that are symmetrically distributed around $\beta = 1$ and share the same value of $|\beta - 1|$ yield identical geometries. Consequently, the band structures are identical for $\beta = 0.7$ and $\beta = 1.3$, $\beta = 0.8$ and $\beta = 1.2$, etc.

As we consider nanoparticles, we employ the dipole approximation. To derive the eigenmodes of the considered system of coupled dipoles, we require the Green's function $\mathds{G}^{\rm tot}(x_i,x_j)$ describing the field at the position of a dipole $i$ sourced by the dipole of particle $j$. By taking all the dipole interactions in the chain into account, we can write the dipole moment of any particle $i$ ($i = 1 \ldots N$) in the chain induced by all other particles as \cite{Messina2013}
\begin{equation}
	\mathbf{p}_i(\omega) = k_0^2 \sum_{j \neq i}^{N} \alpha(\omega) \, \mathds{G}^{\rm tot}(x_i,x_j, \omega) \, \mathbf{p}_j(\omega),
	\label{Eq:Eigenmodes}
\end{equation}
where $\alpha$ is the polarizability of the identical particles, $k_0 = \omega / c$ is the vacuum wave number, $\omega$ is the angular frequency, and $c$ is the speed of light in vacuum. This Green's function $\mathds{G}^{\rm tot}(x_i,x_j, \omega) = \mathds{G}^{0}_{ij} + \mathds{G}^{\rm sc}_{ij}$ can be decomposed into a vacuum part $\mathds{G}^{0}_{ij}$ and a scattering contribution $\mathds{G}^{\rm sc}_{ij}$ which is due to the presence of the substrate. A detailed definition of $\mathds{G}^{0}$ and $\mathds{G}^{\rm sc}$ is given in Ref.~\cite{Naeimi2026}. Here, we choose to align the SSH chain along the x axis parallel to the substrate's surface as sketched in Fig.~\ref{Fig:Sketch}. Note that in the following we will skip the frequency argument for the sake of readibility.

As outlined in Ref.~\cite{Naeimi2026}, the symmetry of the considered system makes it possible to decompose the total Green's function into in-plane (IP) and out-of-plane (OP) contributions
\begin{equation}
	\mathds{G}^{\rm tot}_{ij} = \mathds{G}^{\rm tot}_{ij,\rIP} + \mathds{G}^{\rm tot}_{ij,\rOP},
	\label{Eq:Green}
\end{equation}
for which the IP part characterizes the interacting dipoles in the x-z plane and the OP one, the interaction of dipoles pointing in the y direction. Therefore, the IP Green's function $\mathds{G}^{\rm tot}_{ij,\rIP}$ describes the mixing of the longitudinal modes in the direction along the SSH chain and the transversal modes perpendicular to the substrate, whereas the OP Green's function $\mathds{G}^{\rm tot}_{ij,\rOP}$ describes the transversal modes in the y-direction only. Note that without a substrate, $\mathds{G}^{\rm vac}_{ij,\rIP}$ can be further decomposed into $\mathds{G}^{\rm vac}_{ij,\rIP} = \mathds{G}^{\rm vac}_{ij,\rtrans} + \mathds{G}^{\rm vac}_{ij,\rlong}$ while $\mathds{G}^{\rm vac}_{ij,\rOP} = \mathds{G}^{\rm vac}_{ij,\rtrans}$ holds. This is due to the decoupling of longitudinal and transversal modes because of the vacuum symmetry. 

Eq.~\eqref{Eq:Eigenmodes} describes a set of $3 \times N$ equations which can also be framed in block-matrix notation as
\begin{equation}
	\underline{\mathbf{p}} = k_0^2 \, \alpha \, \underline{\underline{\mathds{G}}} \, \underline{\mathbf{p}},
	\label{Eq:block1}
\end{equation}
where we have introduced the block vector $\underline{\mathbf{p}} = (\mathbf{p}_1, ..., \mathbf{p}_N)^t$ and the block matrix
\begin{equation}
	\underline{\underline{\mathds{G}}}_{ij} = \mathds{G}^{\rm tot}(x_i,x_j) (1 - \delta_{ij}) .
	\label{Eq:blockmatrix}
\end{equation}
Let us stress that for a finite chain of particles, Eq.~\eqref{Eq:block1} can be restated as an eigenvalue equation for the  eigenvectors $\underline{\mathbf{p}}$ such that
\begin{align}
	\underline{\underline{\mathds{M}}} \, \underline{\mathbf{p}} & =  \frac{1}{\alpha} \underline{\mathbf{p}},
	\label{Eq:block2} \\
	\underline{\underline{\mathds{M}}} & = k_0^2 \, \underline{\underline{\mathds{G}}}
\end{align}
holds with eigenvalues $\alpha^{-1}$, which can be used to determine the eigenmodes and eigenfrequencies in a finite chain interacting with the substrate.

However, for an infinite chain of particles we can use the Bloch theorem. Then Eq.~\eqref{Eq:Eigenmodes} can be rewritten as a $2 \times 2$ block matrix equation for the dipole moments of each unit cell for the $\nu= {\rm IP, OP}$ modes as
\begin{equation}
	\mathds{M}^\mathbf{\nu} \begin{pmatrix} p_{A}^{\nu} \\ p_{B}^{\nu} \end{pmatrix} = \frac{1}{\alpha}\begin{pmatrix} p_{A}^{\nu} \\ p_{B}^{\nu} \end{pmatrix}
	\label{Eq:Eigenmodes1}
\end{equation}
with block vectors $p^{\rIP}_{A/B}=(p_{x,A/B},p_{z,A/B})^{t}$, $p^{\rOP}_{A/B} = p_{y,A/B}$ and block matrix \cite{Ford2004}
\begin{equation}
	\mathds{M}^\mathbf{\nu} = \begin{pmatrix} M_{AA}^{\nu} & M_{AB}^{\nu} \\ M_{BA}^{\nu} & M_{BB}^{\nu} \end{pmatrix}
\end{equation}
where 
\begin{align}
	M_{AA}^{\nu} & =  M_{BB}^{\nu} = k_0^2  \sum_{j \in \mathds{Z}, j \neq 0} \mathds{G}_\nu (jd) \re^{\ri k_x j d},  
	\label{Eq:MAA} \\
	M_{AB/BA}^{\nu} &=  k_0^2 \sum_{j \in \mathds{Z}} \mathds{G}_\nu (jd \pm t) \re^{\ri k_x j d}.
	\label{Eq:MAB}
\end{align}
Note that $\mathds{G}_\nu$ replaces $\mathds{G}_{ij}^{\rm tot}$ in Eq.~\eqref{Eq:Green} with the spatial arguments $x_i-x_j = j d$ if the particles $i$ and $j$ belong to the same sublattice and $x_i - x_j = j d \pm t$ if they belong to different sublattices. Herein, the eigenvectors are $(p_{A}^{\nu}, p_{B}^{\nu})^t$ for the coupled IP and OP modes from which the eigenfrequencies for each value of $k_x$, i.e. the frequency bands, can be derived. 

As a quantitative measure to determine the topological phase, we define the Zak phase which serves as topological invariant as follows~\cite{Pocock2018, Asboth2016}
\begin{equation}
	\tilde{\gamma}_\nu = \ri \int_{0}^{2 \pi/d} \rd k_x\, \left( \mathbf{p}_{L}^{\nu}\right)^{\dagger} \cdot \left(\pdv{\mathbf{p}_{R}^{\nu}}{k_x}\right).
	\label{Eq:Zak}
\end{equation}
Here, $\mathbf{p}_{L/R}^{\nu}$ are the normalized biorthogonal left and right eigenvectors of Eq.~\eqref{Eq:Eigenmodes1}. The calculated Zak phase is defined modulo $2 \pi$ as $\tilde{\gamma} = \gamma \pm 2 n \pi$ for $n \in \mathds{Z}$ so that the value $\gamma = \pi$ describes a topological non-trivial phase and $\gamma = 0$ describes the trivial phase~\cite{Asboth2016}. The bulk-edge correspondence then predicts that in the topological non-trivial phase there exist edge modes for finite chains. Note that this quantization of the biorthogonal Zak phase is protected by the spatial inversion symmetry of the Bloch matrix $\mathds{M}$. By introducing the parity operator $\mathcal{P}_\nu = \sigma_x \otimes \mathds{1}_\nu$ where $\sigma_x$ is the corresponding Pauli matrix and $\mathds{1}_\nu$ is the identity matrix for the specified polarization $\nu$, it can be shown that
\begin{equation}
	\mathcal{P}_\nu \mathds{M}^\nu (k_x) \mathcal{P}_\nu^{-1} = \mathds{M}^\nu (-k_x).
	\label{Eq:parity}
\end{equation}
This follows directly from the symmetry of the Green's functions \cite{Naeimi2026}, whose $k_x$-space components are stricly even in $k_x$, with the exception of the xz/zx scattering terms in $\mathds{G}_{ij}^{\rm sc}$, which are odd in $k_x$. Together with Lorentz reciprocity, which satisfies $\mathds{M}^\nu(-k_x) = \mathds{M}^{\nu, t}(k_x)$, this rigorously ensures that the biorthogonal Zak phase is strictly quantized to $0$ or $\pi$.

To support infrared surface plasmons, we use InSb for both the nanoparticles and the substrate, modeled by the Drude permittivity~\cite{Law2014}
\begin{equation}
	\varepsilon = \varepsilon_\infty \left(1 - \frac{\omega_{\rm p}^2}{\omega(\omega+{\rm i}\Gamma)} \right).
	\label{eps1}
\end{equation}
Accounting only for the electronic part of the optical response, we set $\epsilon_\infty = 15.68$, $\Gamma = 1\times10^{12}$ rad/s, and the effective mass $m^* = 7.29\times10^{-32}$ kg. The plasma frequency of the free carriers $\omega_{\rm p} = (\frac{ne^2}{m^*\varepsilon_0\varepsilon_\infty})^{\frac{1}{2}}$ is determined by the charge carrier concentration, fixed at $n = n_{\rm p} = 1.36\times10^{19}$ cm$^{-3}$ for the nanoparticles and varied between $n_{\rm sub} = \{1.34, 1.36\} \times10^{19}$ cm$^{-3}$ for the substrate.

In the quasistatic limit ($R k_0 \ll 1$) and within the dipole approximation, the particle polarizability is
\begin{equation}
	\alpha = 4\pi R^3\frac{\varepsilon-1}{\varepsilon + 2}
\end{equation}
yielding a localized plasmon (LP) resonance at $\omega_{\rm LP} = \omega_p \sqrt{\epsilon_\infty/(\epsilon_\infty + 2)} = 1.752\times10^{14}$ rad/s. For the planar InSb substrate, the surface modes satisfy the dispersion relation
\begin{equation}
	k_\perp^{\rm SPP} = k_{0}\sqrt{\frac{\epsilon_{\rm sub}(\omega)}{\epsilon_{\rm sub}(\omega) + 1}}
	\label{Eq:DispSPP}
\end{equation}
where $k_\perp^{\rm SPP}$ is the wave vector of the surface wave. The surface mode resonance frequency is defined by $\Re[\epsilon_{\rm sub}(\omega_{\rm SPP})] = -1$ and $\Im[\epsilon_{\rm sub}(\omega_{\rm SPP})] \ll 1$. Since $\omega_{\rm SPP} > \omega_{\rm LP}$ for all considered doping levels of the substrate, the localized particle resonances can consistently couple to the substrate's surface modes.

%
%
\section{Band structure and Edge Modes}\label{sec:band}

The topological phase and the resulting band structure can be modulated by three primary parameters: the dimerization parameter $\beta$, the vacuum gap $z$ between the substrate and the chain, and the substrate's charge carrier density $n_{\rm sub}$. As established in Ref.~\cite{Naeimi2026}, $n_{\rm sub}$ directly controls the coupling strength between the chain and the substrate by shifting the SPP resonance frequency $\omega_{\rm SPP}$ relative to the LP resonance $\omega_{\rm LP}$ of the particles. This interaction induces a hybridization of the dipole-active bands -- specifically the upper transverse (z) and the lower longitudinal (x) branches. To complement this, in the following, we analyze the emergence of edge modes as a function of $\beta$ and $z$ to discuss the possible control of the band structure and the topological phase transition by means of these external parameters. In addition to the chosen parameters, other variables also influence the band structure. Varying the doping level $n_p$ of the particles alters the resonance frequency of the nanoparticles, thereby primarily shifting the bands in frequency. Additionally, the damping constant $\Gamma$ determines the linewidth of the resonances in both the substrate and the nanoparticle.

\subsection{Dependence on the dimerization}\label{sec:beta}

\begin{figure*}
	\centering
	\includegraphics[width=\textwidth]{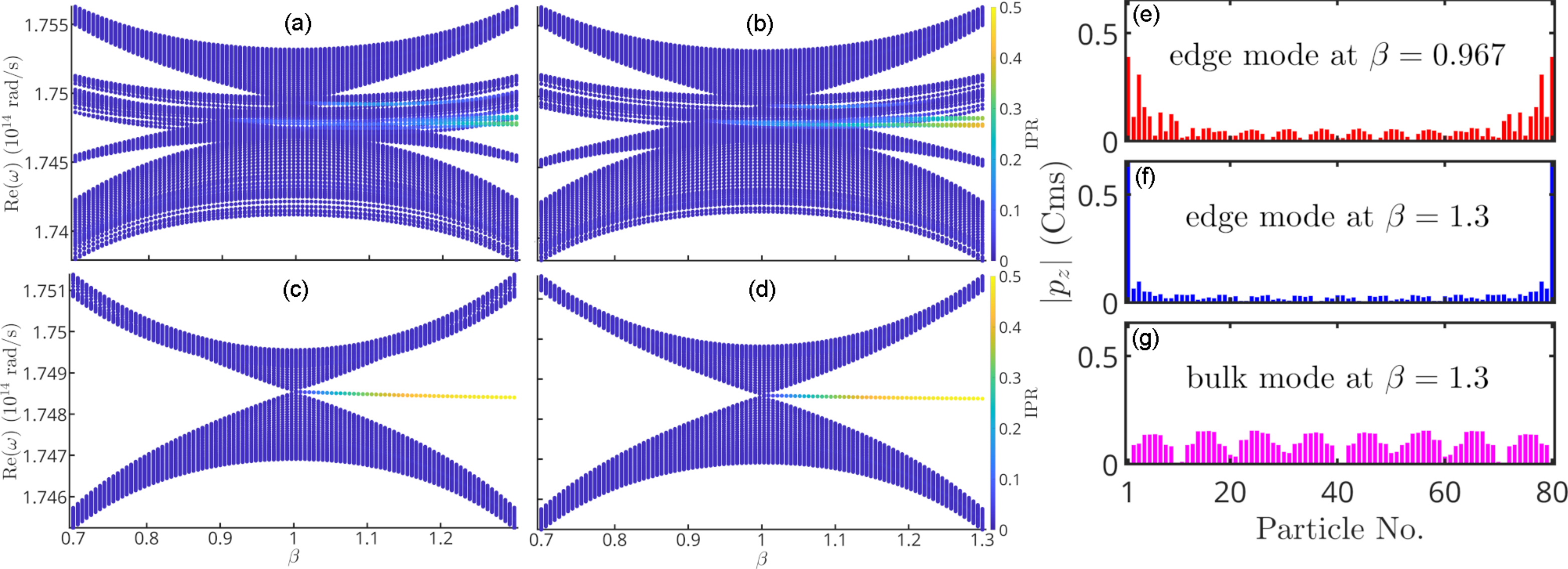}
	\caption{Eigenfrequencies and IPR values for IP (a),(b) and OP modes (c),(d) for $n_{\rm sub} = 1.34 \times 10^{19}$ cm$^{-3}$ (a),(c) and $n_{\rm sub} = 1.36 \times 10^{19}$ cm$^{-3}$ (b),(d) at $z=500$ for different $\beta$ and a total of $N=80$ particles. The localization of selected modes from (b) is shown by means of the dipole moments for $\omega = 1.7483 \times 10^{14}$ rad/s (e), $\omega = 1.7480 \times 10^{14}$ rad/s (f), and $\omega = 1.7452 \times 10^{14}$ rad/s (g).}
	\label{Fig:dots_beta}
\end{figure*}
We begin by examining the band structure of the finite SSH chain with $n_{\rm sub} = \{ 1.34, 1.36 \} \times 10^{19}$ cm$^{-3}$, $N=80$ particles, and $z = 500$ nm which is depicted in Fig.~\ref{Fig:dots_beta}. To quantify the localization of each eigenmode, we utilize the Inverse Participation Ratio (IPR) \cite{Gong2024}
\begin{equation}
	{\rm IPR} = \frac{\sum_{i=1}^N |\mathbf{p}i|^4}{\left( \sum_{i=1}^N |\mathbf{p}_i|^2 \right)^2}.
\end{equation}
High IPR values indicate a strong spatial confinement of the dipole moments. For edge modes, which localize on both ends of the chain, a maximum IPR value of $\text{IPR}=0.5$ could be reached, while completely delocalized eigenmodes have $\text{IPR}=1/N$. Therefore, this quantifier is very useful for identifying localized edge modes, as can be seen in Fig.~\ref{Fig:dots_beta}. 

Figs.~\ref{Fig:dots_beta}(c),(d) show the two OP bands for the transversal modes in the y direction as a function of $\beta$. For the topological trivial phase $\beta < 1$, there are no modes in the band gap. For $\beta=1$, the band gap closes and then reopens for $\beta > 1$. It can be clearly seen that for $\beta = 1$, two edge modes in the middle of the band gap emerge which indicates the topological phase transition at $\beta = 1$. The localization of these edge modes increases with increasing $\beta$ as indicated by the IPR. This behavior remains unchanged when changing the doping $n_\text{sub}$ of the substrate. 

For the IP modes, we clearly observe four edge modes at $\beta=1.3$ and $n_\text{sub}=1.36 \times 10^{19}$ cm$^{-3}$ in Fig.~\ref{Fig:dots_beta}(b). The upper ``edge mode branch'' is x polarized while the lower one is z polarized. This information can be gathered from the eigen dipole moments (not shown). Note that two edge modes always have almost identical frequencies and, therefore, overlap in Fig.~\ref{Fig:dots_beta}. However, when reducing $\beta < 1.2$, a second ``edge mode branch'' at higher frequencies emerges while one of the low frequency branches fades out. This new edge mode branch remains intact while reducing $\beta$ to $\beta=1$. The lower remaining edge mode branch splits, so that the previously overlapping two edge modes separate in frequency and become distinguishable. This branch even remains for values $0.9<\beta<1$, which is surprising because for the SSH-chain without a substrate the phase transition is expected to happen at $\beta = 1$. To illustrate the localization of this mode at the chain edges, we show the dipole moments along the chain for $\beta=0.967$ at the frequency corresponding to a pronounced IPR value in Fig.~\ref{Fig:dots_beta}(e), and compare it with an edge mode in Fig.~\ref{Fig:dots_beta}(f) and a bulk mode at $\beta=1.3$ in Fig.~\ref{Fig:dots_beta}(g). Notably, while the dipole moments at the edge mode frequency are less localized at the chain ends for $\beta=0.967$ than for $\beta=1.3$, they still clearly reveal a much stronger decay from the edges compared to the standing wave pattern observed for bulk modes. In Fig.~\ref{Fig:dots_beta}(a), we observe the same behavior but the emergence of the upper edge mode branch starts already for $\beta<1.25$. This tendency of a shift to larger $\beta$ values continues for smaller substrate dopings (not shown). Therefore when $\omega_{\rm SPP}$ approaches $\omega_{\rm LP}$, the upper edge mode branch becomes more important than the lower ones, and vice versa.

\begin{figure}
	\centering
	\includegraphics[width=0.5\textwidth]{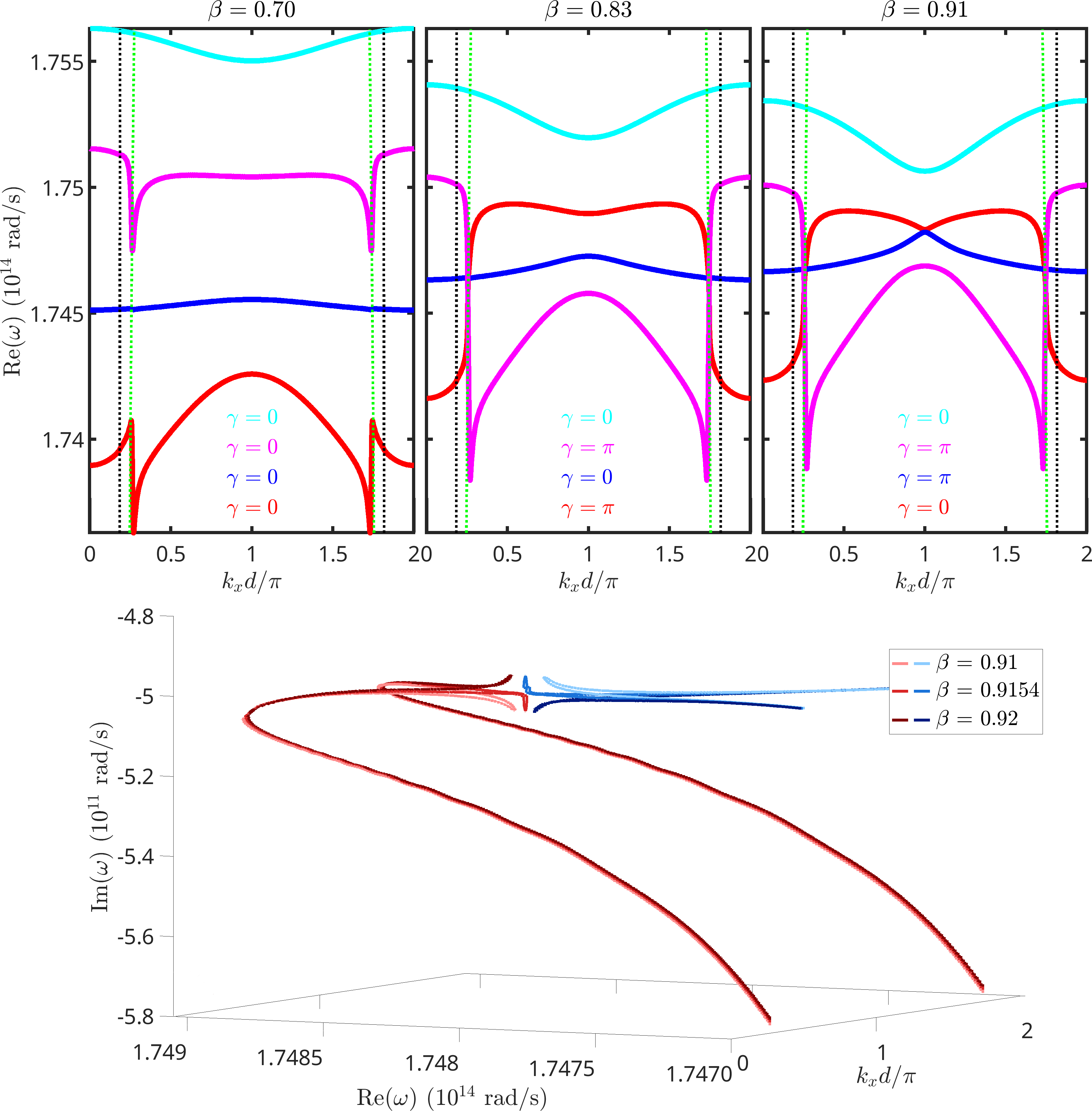}
	\caption{Upper panel: IP band structure for $z=500$ nm and $n_\text{sub} = 1.36 \times 10^{19}$ cm$^{-3}$ for different $\beta$ values. For each band, the Zak phases are given in the corresponding colors. The dashed lines correspond to the light lines in vacuum (black) and the surface polariton light line, i.e.\ the surface mode dispersion relation, of the InSb substrate (green) taken from Eq.~\eqref{Eq:DispSPP}. Lower panel: 3D illustration $(k_x, \text{Re}(\omega), \text{Im}(\omega))$ of the red and blue bands for $\beta$ values close to the band touching.}
	\label{Fig:IP_bands_beta}
\end{figure}

To understand the mechanisms behind the evolution of different edge mode branches in more detail, we show in Fig.~\ref{Fig:IP_bands_beta} the band structure of a corresponding infinite SSH chain and the calculated Zak phases taken from Eq.\eqref{Eq:Zak}. Note that the behavior of the bands is symmetric with respect to $\beta=1$. For dimerizations $\beta<0.83$, all four bands are separated but due to the coupling with the substrate, we observe a strong variation of the dipole-active bands (red and magenta) close to the surface plasmon polariton light line of the InSb substrate (dashed green), indicating strong interaction of the dipole-active chain modes with the surface modes. At this stage, no band has acquired a Zak phase so that the chain is clearly in a trivial phase. In the following, we identify two mechanisms which explain the behavior that we found for the edge mode branches observed in Fig.~\ref{Fig:dots_beta}(b) by increasing $\beta$.

\subsubsection{Mechanism 1: Band hybridization and inversion}

As we continue to increase $\beta$, the coupling between the two dipole-active bands and the surface modes increases close to the surface modes' dispersion relation. At $\beta=0.83$, the bands touch and swap. Therefore we find for the intermediate parts of the dipole-active bands a band inversion. This band inversion remains up to the value $\beta=1.17$ due to the symmetry with respect to $\beta=1$. It can be seen that due to the band inversion, both dipole-active bands acquire a Zak phase of $\gamma=\pi$. While the single-band Zak phase serves as a well-defined topological invariant predicting edge-mode existence for isolated spectral gaps, the non-Abelian Wilson loop reveals that the composite-band Berry phase of the coupled subspace vanishes (modulo $2 \pi$) in the strongly hybridized regime. This confirms that individual-band phase shifts within the hybridized structure are gauge-dependent artifacts of band-mixing rather than genuine gap invariants. Since this hybridization happens at the InSb surface mode light line, one can explain this feature by the influence of the substrate's SPPs. 

\begin{figure}
	\centering
	\includegraphics[width=0.5\textwidth]{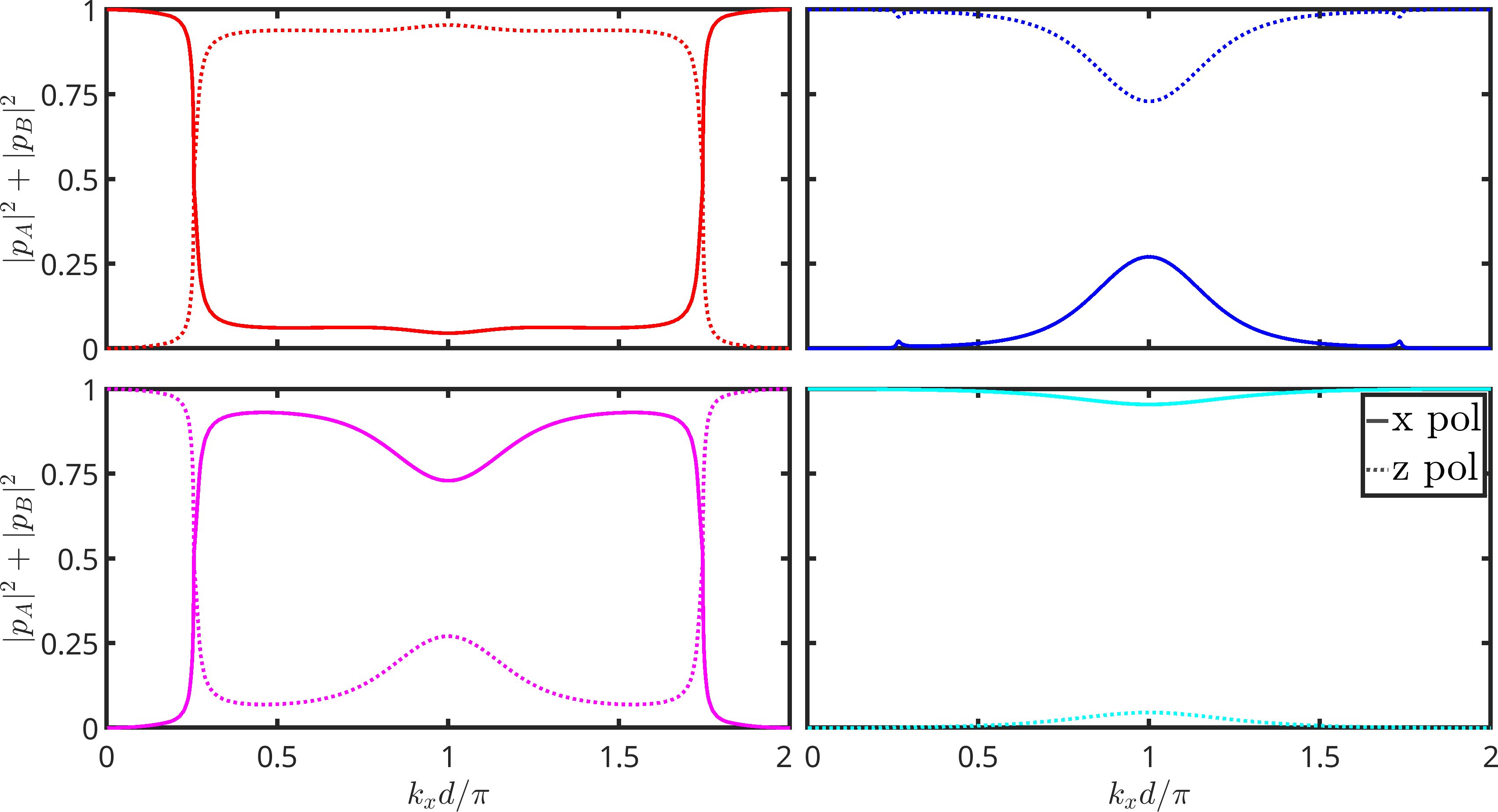}
	\caption{Polarizations of the unit cell $|p_A|^2+|p_B|^2$ in x (solid) and z (dashed) direction for each band in Fig.~\ref{Fig:IP_bands_beta} at $\beta=0.83$.}
	\label{Fig:IP_Pol1}
\end{figure}

In Fig.~\ref{Fig:IP_Pol1}, we show the polarizations of the unit cell at $\beta=0.83$ for each band in Fig.~\ref{Fig:IP_bands_beta}. The red and cyan bands are longitudinal IP modes, and therefore, they are typically x-polarized. The magenta and blue bands refer to transversal IP modes, and are, therefore, typically z-polarized. Due to the hybridization and band inversion at the intersection with the substrate's surface polariton light line, the polarization of the modes flips, which corresponds to the partial swap of the red and magenta bands. This is an indication of the strong interaction of the longitudinal and transversal IP modes due to the interaction via the surface modes of the substrate. Therefore, the resulting bands are now clearly hybrid longitudinal-transversal IP modes.

\begin{figure}
	\centering
	\includegraphics[width=0.5\textwidth]{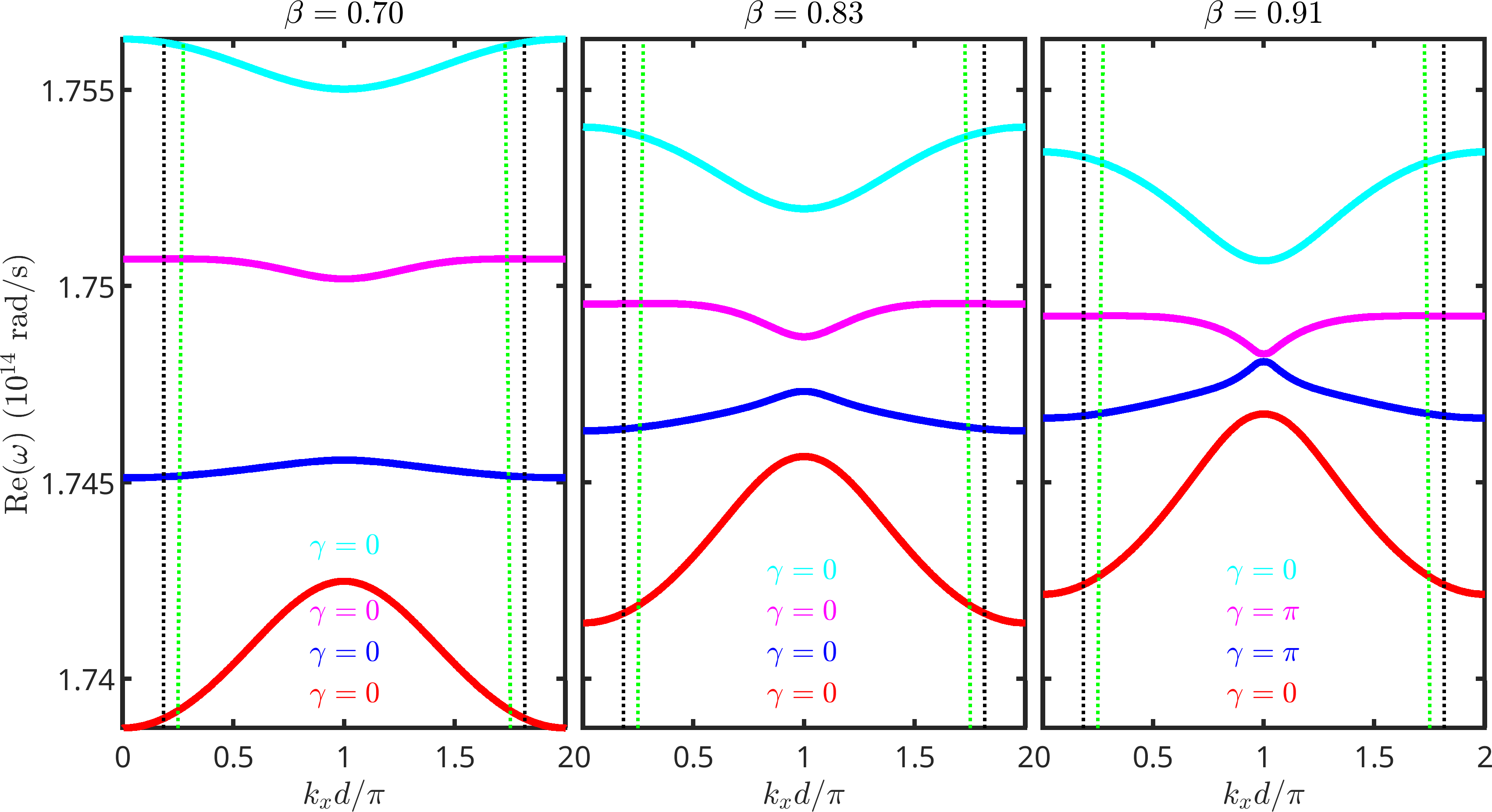}
	\caption{Identical IP band structure to Fig.~\ref{Fig:IP_bands_beta} but with the nearest neighbor approximation.}
	\label{Fig:IP_bands_beta_nn}
\end{figure}

To fully understand this mechanism, we compare the results in Fig.~\ref{Fig:IP_bands_beta} with the ones obtained in the nearest-neighbour approximation which are shown in Fig.~\ref{Fig:IP_bands_beta_nn}. By again choosing $\beta=0.83$, we find that the hybridization between the dipole-active bands has disappeared. Therefore the hybridization and nearly complete inversion of the mode character (longitudinal or transversal) close to the surface polariton light line can be associated with the long-range interactions along the chain mediated via the substrate's SPPs. A quanitative analysis also shows that this hybridization starts to appear only after allowing at least 77 neighboring interactions.

At $\beta=1.17$, where the band structures are identical to $\beta=0.83$ for infinite chains, we observed in Fig.~\ref{Fig:dots_beta}(b) the emergence of the upper edge mode branch so that we assign this emergence to the hybridization of the two dipole-active bands accompanied by an individual non-vanishing Zak phase. We find that all the bands have an individual non-vanishing Zak phase for $1.17<\beta<1.3$, and the two dipole-active bands acquire an individual Zak phase of zero during the hybridization for $\beta < 1.17$. Interestingly, however, we do not observe an emergence of edge mode branches for $\beta=0.83$ even though the bands acquire a Zak phase. Hence, this mechanism only plays a role in the topologically non-trivial phase.

\subsubsection{Mechanism 2: Band touching and gap reopening}

In Fig.~\ref{Fig:dots_beta}(b), we clearly identify emerging edge modes for $\beta\approx0.91$. In Fig.~\ref{Fig:IP_bands_beta}, this corresponds to a touching of the lower transversal band (blue) and the upper hybridized band (red). There is obviously a closing and reopening of the gap between these two bands, by which they acquire another Zak phase $\gamma=\pi$, so that the resulting Zak phase of the hybridized band (red) is zero and that of the transversal band (blue) is $\pi$. This mechanism is identical to the one that is causing the topological phase transition in SSH chains at $\beta=1$ (as clearly observed for the OP modes). The lower panel of Fig.~\ref{Fig:IP_bands_beta} illustrates this behavior via a 3D plot in the complex frequency plane. At $\beta=0.9154$, the two bands touch at two points very close to the center of the Brillouin zone (indicated in medium red and medium blue). For larger and smaller values of $\beta$, the bands separate. The reopened band gap makes room for topological edge modes, which explains why only this mechanism leads to visible edge modes in Fig.~\ref{Fig:dots_beta}(b) for $\beta<1$. On the other hand the hybridization does not open a band gap, and therefore, this mechanism is not generating edge modes. Although both mechanisms lead to an acquisition of a Zak phase of $\gamma=\pi$. At $\beta = 1.09$, which is the symmetric value to  $\beta=0.91$, one can see in  Fig.~\ref{Fig:dots_beta}(b) a dip in the IPR value in the lower branches, which indicates this transition for $\beta > 1$.

\begin{figure}
	\centering
	\includegraphics[width=0.5\textwidth]{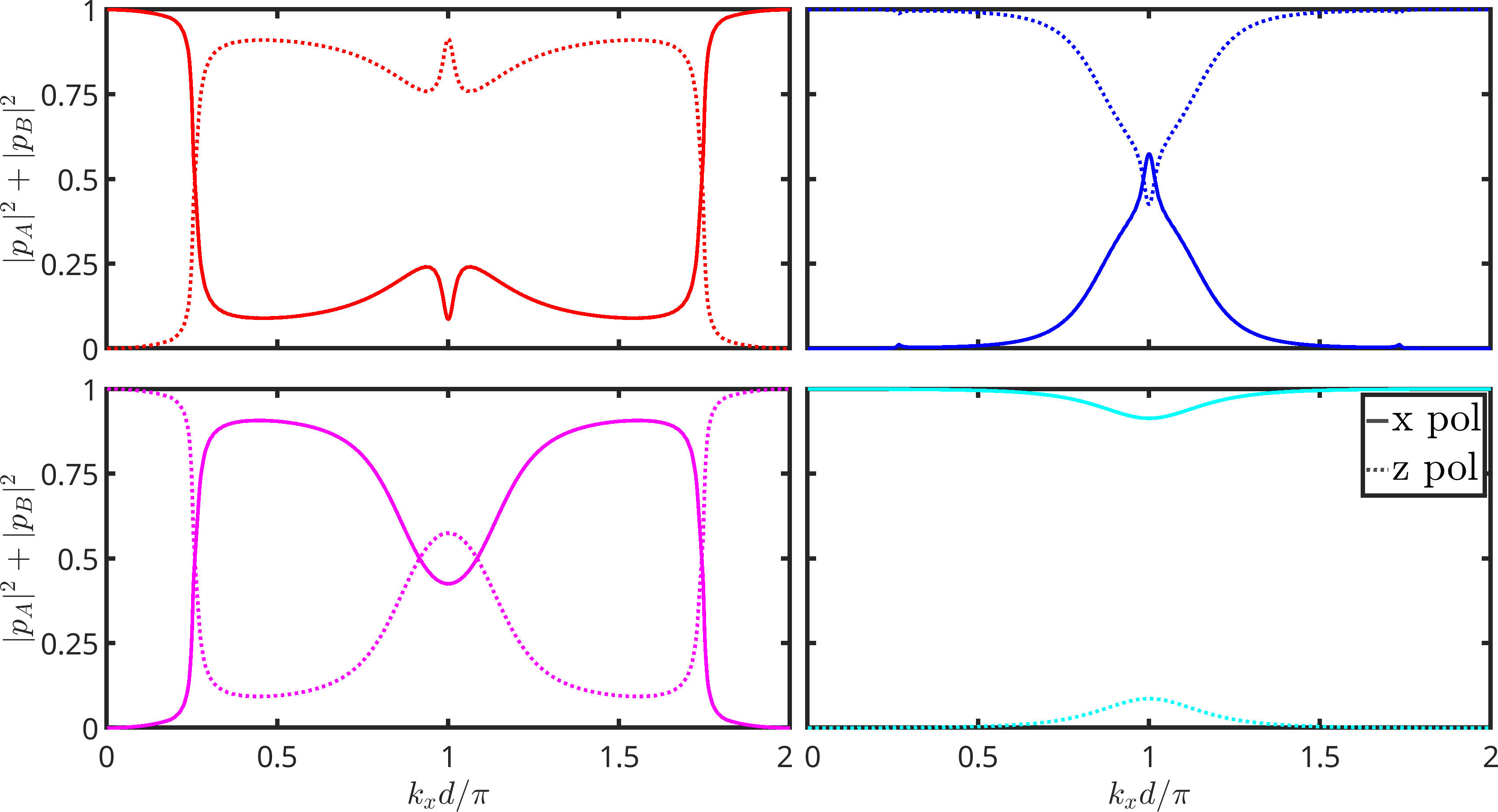}
	\caption{Polarizations of the unit cell $|p_A|^2+|p_B|^2$ in x (solid) and z (dashed) direction for each band in Fig.~\ref{Fig:IP_bands_beta} at $\beta=0.91$.}
	\label{Fig:IP_Pol2}
\end{figure}

\begin{figure}
	\centering
	\includegraphics[width=0.5\textwidth]{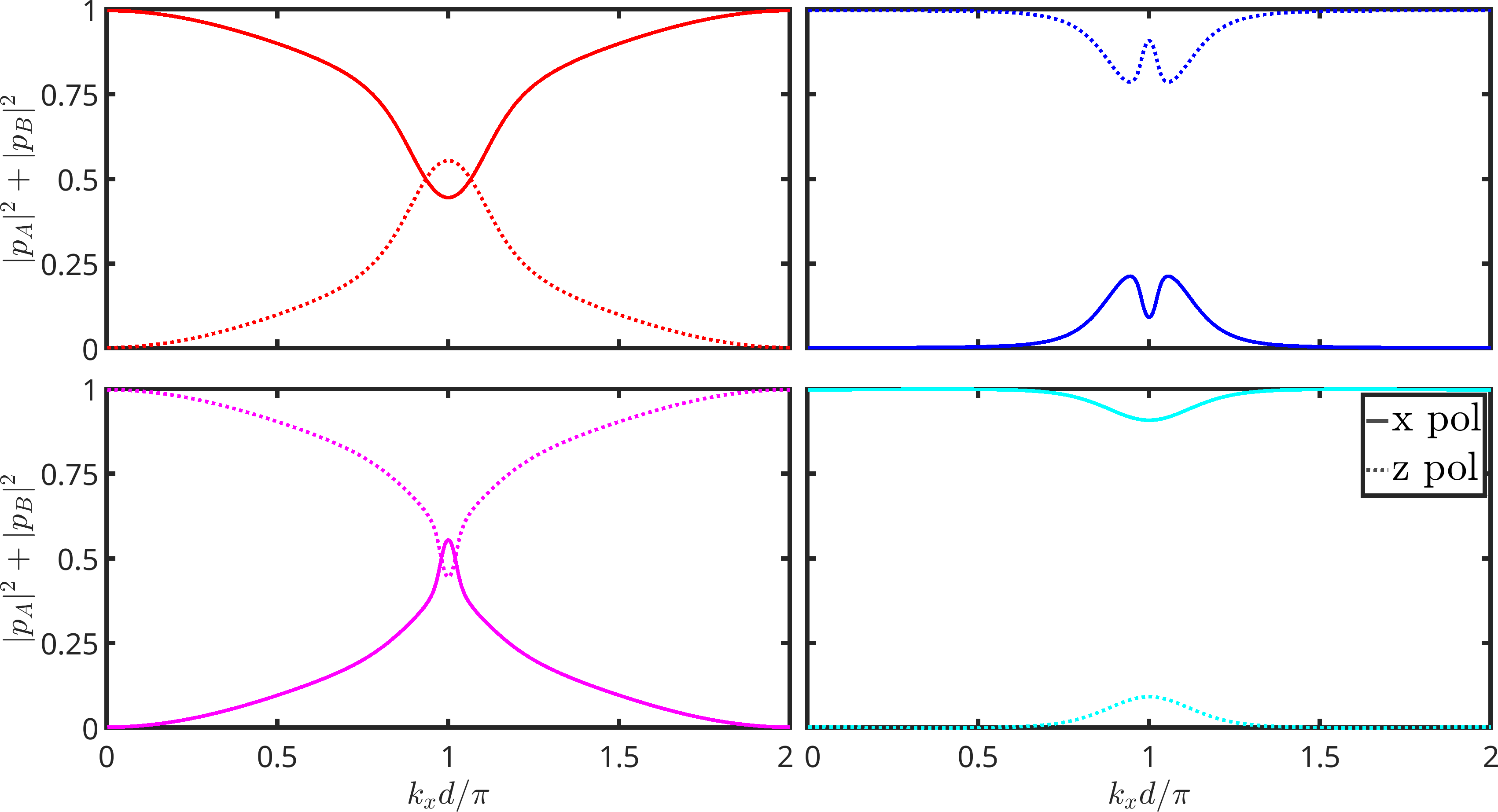}
	\caption{Polarizations of the unit cell $|p_A|^2+|p_B|^2$ in x (solid) and z (dashed) direction for each band in Fig.~\ref{Fig:IP_bands_beta_nn} at $\beta=0.91$ in nearest-neighbor approximation.}
	\label{Fig:IP_Pol3}
\end{figure}

For $\beta=0.91$, we show in Fig.~\ref{Fig:IP_Pol2} again the polarization. We find an exchange of the polarization between the two bands at the center of the Brillouin zone where the red and blue bands touch. The peak visible for the x polarization of the red band in the center of the Brillouin zone corresponds to a dip in the x polarization of the blue one and vice versa for the z polarization, which acts like a superposition. Meanwhile, the hybridization is still visible, and the magenta bands also flip from x to z polarization, which seems unrelated to the touching of the red and blue bands. This helps to distinguish these two mechanisms such that a band touching leads to an addition or subtraction of polarization at the center of the Brillouin zone while a hybridization leads to a complete flipping of the polarization. When only accounting for nearest-neighbor interactions we still find this superposition for the polarization in Fig.~\ref{Fig:IP_Pol3} (corresponding bands are shown in Figs.~\ref{Fig:IP_bands_beta_nn}), but for the blue and magenta bands instead, due to the missing hybridization. It can be concluded that the band touching mechanism is a short-range effect mediated by the substrate. On the other hand the hybridization mechanism is not visible in the polarization for nearest-neighbor interaction and, therefore, is a pure long-range effect.  

\subsection{Dependence on the chain-substrate distance}\label{sec:z}

\begin{figure*}
	\centering
	\includegraphics[width=\textwidth]{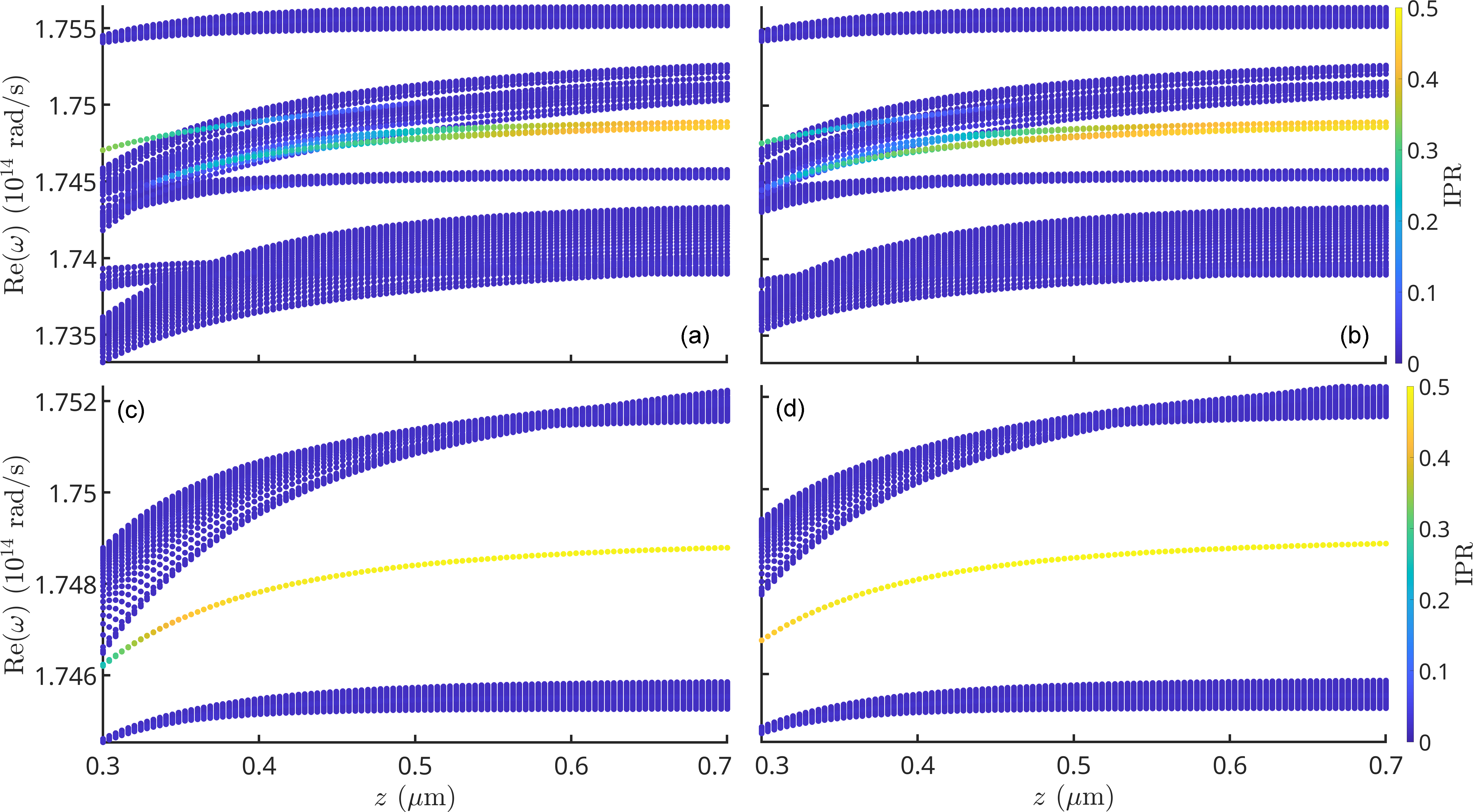}
	\caption{Eigenfrequencies and IPR values for IP (a),(b) and OP modes (c),(d) for $n_{\rm sub} = 1.34 \times 10^{19}$ cm$^{-3}$ (a),(c) and $n_{\rm sub} = 1.36 \times 10^{19}$ cm$^{-3}$ (b),(d) at $\beta=1.3$ for different $z$.}
	\label{Fig:dots_z}
\end{figure*}

After understanding the main mechanisms when varying the dimerization $\beta$, we now keep $\beta=1.3$ and vary $z$ instead. Since this effectively varies the coupling between the chain and the substrate's surface modes, which is the reason for the observed mechanisms, we can expect to find similar mechanisms as for the variation of $\beta$. As depicted in Fig.~\ref{Fig:dots_z}(c),(d), there is for any distance always a band gap between the two OP bands ensuring an edge mode for all considered gaps sizes $z$ between the chain and substrate in the non-trivial phase. For smaller doping, such as $n_\text{sub}=1.34 \times 10^{19}$ cm$^{-3}$, only a slight redshift for the eigenmodes at smaller gap sizes is visible compared to $n_\text{sub}=1.36 \times 10^{19}$ cm$^{-3}$.

The IP spectrum in Fig.~\ref{Fig:dots_z}(b) shows the same mechanism as for constant $z$ but varying $\beta$. Again, we have two close edge mode branches for large $z$. For $n_\text{sub}=1.34 \times 10^{19}$ cm$^{-3}$, this difference increases for decreasing gap sizes and around $z=410$ nm the upper edge mode branch fades out and is replaced by another one at higher frequencies. The localization of the new upper edge mode branch increases for decreasing $z$ while the remaining lower edge mode branch finally vanishes for small $z$, which happens around $z=310$ nm. For smaller $n_\text{sub}$, both effects start already for larger $z$, reflecting a stronger coupling between the chain and substrate due to a smaller frequency mismatch between the resonance frequencies of the substrate and individual chain particles. 

\begin{figure}
	\centering
	\includegraphics[width=0.5\textwidth]{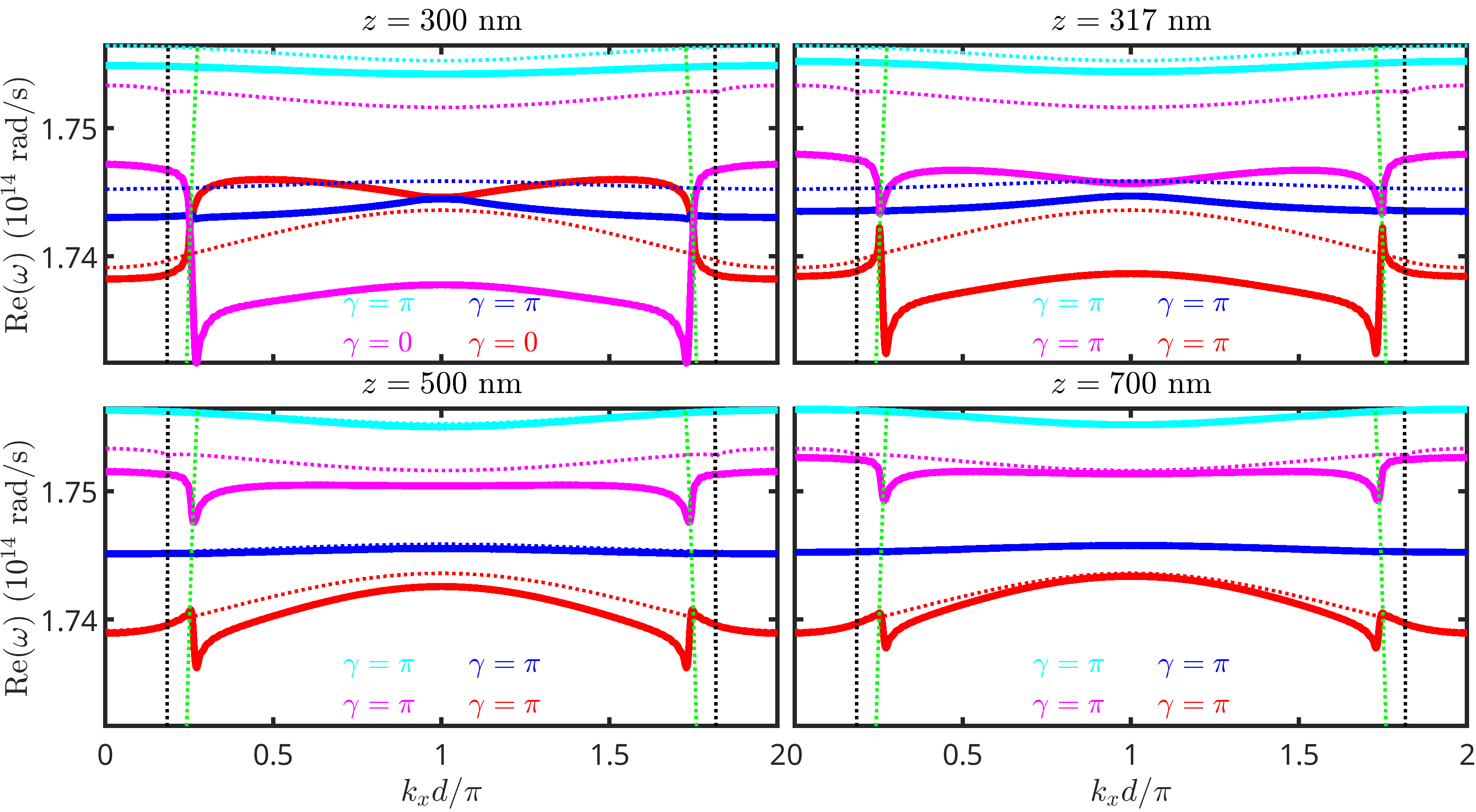}
	\caption{IP band structure for $\beta=1.3$ and $n_\text{sub} = 1.36 \times 10^{19}$ cm$^{-3}$ for different $z$ values. For each band, the Zak phases are given in the corresponding colors. The dashed lines correspond to the light lines in vacuum (black) and in the InSb substrate (green) taken from Eq.~\eqref{Eq:DispSPP} as well as the bands for a chain in vacuum (same colors as the bands).}
	\label{Fig:IP_bands_z}
\end{figure}

In Fig.~\ref{Fig:IP_bands_z}, we show the bands of an infinite chain. One can clearly see similar mechanisms happening as when varying $\beta$. For $z>317$ nm, there is no hybridization between the dipole-active bands, and there are clearly visible band gaps between all bands, each having a Zak phase of $\gamma=\pi$. For $300<z<317$ nm the hybridization between the dipole-active bands adds a Zak phase of $\gamma=\pi$ to the involved red and magenta bands, so that in total their Zak phases vanish. This, again, coincides with the fading of the upper edge mode branch in Fig.~\ref{Fig:dots_z}(b). For reference, we also show the results for free-standing chains without a substrate corresponding to $z \rightarrow \infty$ as dashed lines in Fig.~\ref{Fig:IP_bands_z}. As expected, the bands for finite $z$ approach this limit of large $z \rightarrow \infty$ for increasing chain substrate distances due to a weaker interaction between the substrate and chain resonances. For the dipole-inactive bands (cyan and blue), this asymptotic behavior happens faster than for the dipole-active bands.

To briefly summarize at this stage, the variation of $\beta$ and $z$ has shown two main mechanisms that play a decisive role in the emergence and vanishing of Zak phases and edge modes: band hybridization and band touching. Band touching is clearly visible when comparing the IPR values, regardless of $\beta>1$ or $\beta<1$, and can be only found in the center of the Brillouin zone. Band hybridizations, however, seem to be only visible for $\beta>1$ by a fading in the IPR values due to a loss of individual Zak phase and can be identified as a band swapping for infinite bands at the intersections with the InSb substrate's surface polariton light line.

\subsection{Robustness against disorder}\label{sec:robust}

To demonstrate the robustness of the topologically protected edge modes against structural disorder, we introduce a random displacement to the position of each particle in the chain. The displacement is drawn from a uniform distribution within the range $-x_{r} \leq x \leq x_{r}$, with the maximum deviation defined as $x_r = p t$. In our analysis, we consider perturbation strengths of $p = 7\%$ such that $t_{min} > 3 R$ is always fulfilled to ensure the validity of the dipole approximation, while fixing the displacement scale for this calculation at $\beta = 0.7$ to ensure a consistent magnitude of disorder across all $\beta$ values. 

\begin{figure*}
	\centering
	\includegraphics[width=\textwidth]{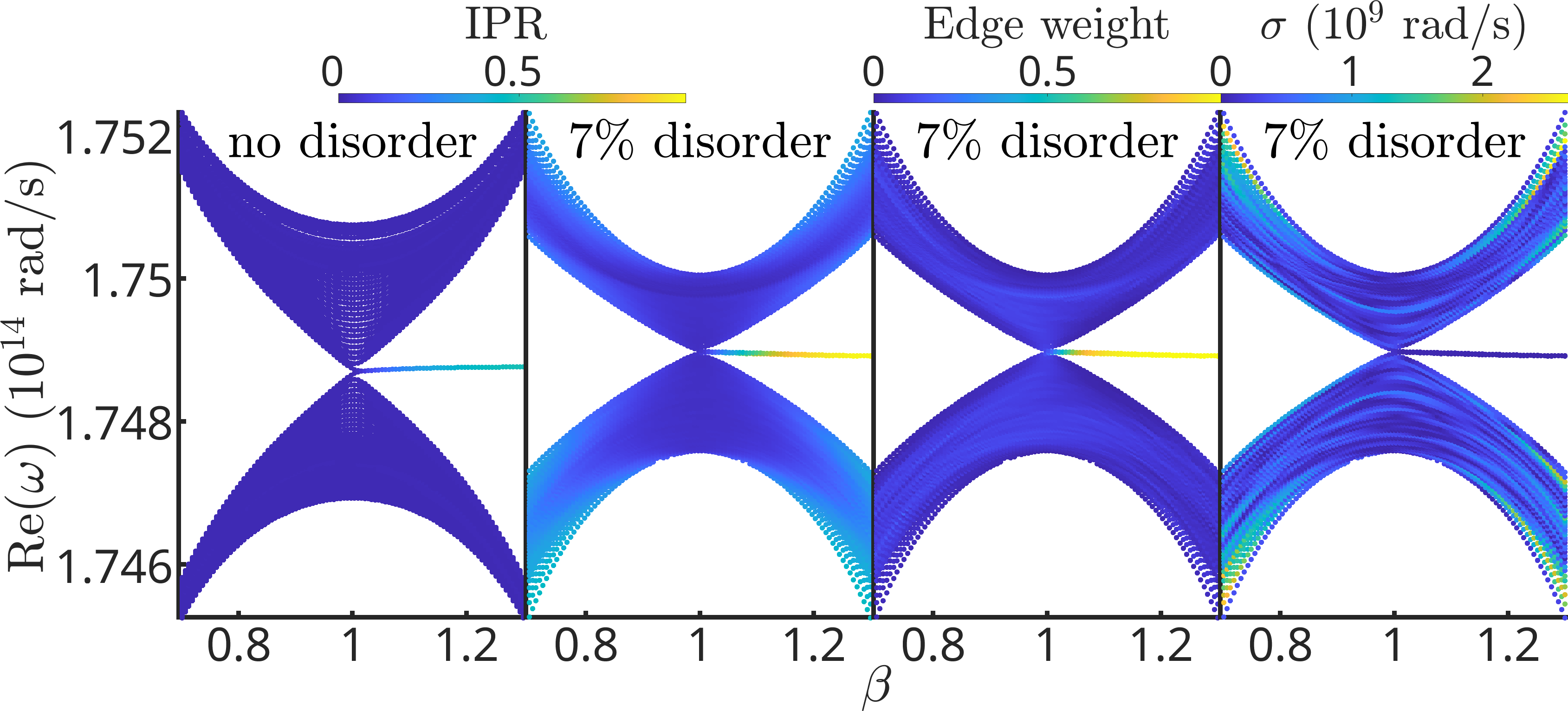}
	\caption{Robustness of the edge modes against disorder. The OP bands of Fig.~~\ref{Fig:dots_beta}(d) are shown for $p = 7 \%$ disorder in terms of IPR values, edge weights, and standard deviations averaged over 100 configurations.}
	\label{Fig:robust_OP}
\end{figure*}

Fig.~\ref{Fig:robust_OP} illustrates the resulting OP eigenfrequencies for $n = 1.36 \times 10^{19}$ cm$^{-3}$ and $z=500$ nm compared with the averaged eigenfrequencies for 100 perturbed configurations. When comparing the unperturbed and perturbed cases, the edge modes are still clearly visible. The disorder even increases their localization. Furthermore, disorder makes the bulk bands broader and increases the IPR values for all dimerizations $\beta$ regardless of the topological phase. This is to be expected, as disorder introduces defects, and therefore localized defect modes. This can be demonstrated by comparing the IPR values with the standard deviation $\sigma$ and the edge weights, which are defined as
\begin{equation}
	{\rm edge \, weight} = \frac{\sum_{i \in \text{edges}} |\mathbf{p}i|^2}{\sum_{i=1}^N |\mathbf{p}_i|^2}
\end{equation}
and quantify localization solely at the chain ends. For this definition, we consider the dipole moments of the first and last two unit cells. The edge weights reveal the strong edge localization expected for topologically protected edge modes, whereas the IPR values also yield high values for localized defects resulting from particle isolation within the chain. For the OP eigenfrequencies, defect localization seems to be negligible. The standard deviation confirms the robustness of the edge modes against disorder, as it yields significantly smaller values for the edge modes than for the bulk modes.

\begin{figure*}
	\centering
	\includegraphics[width=\textwidth]{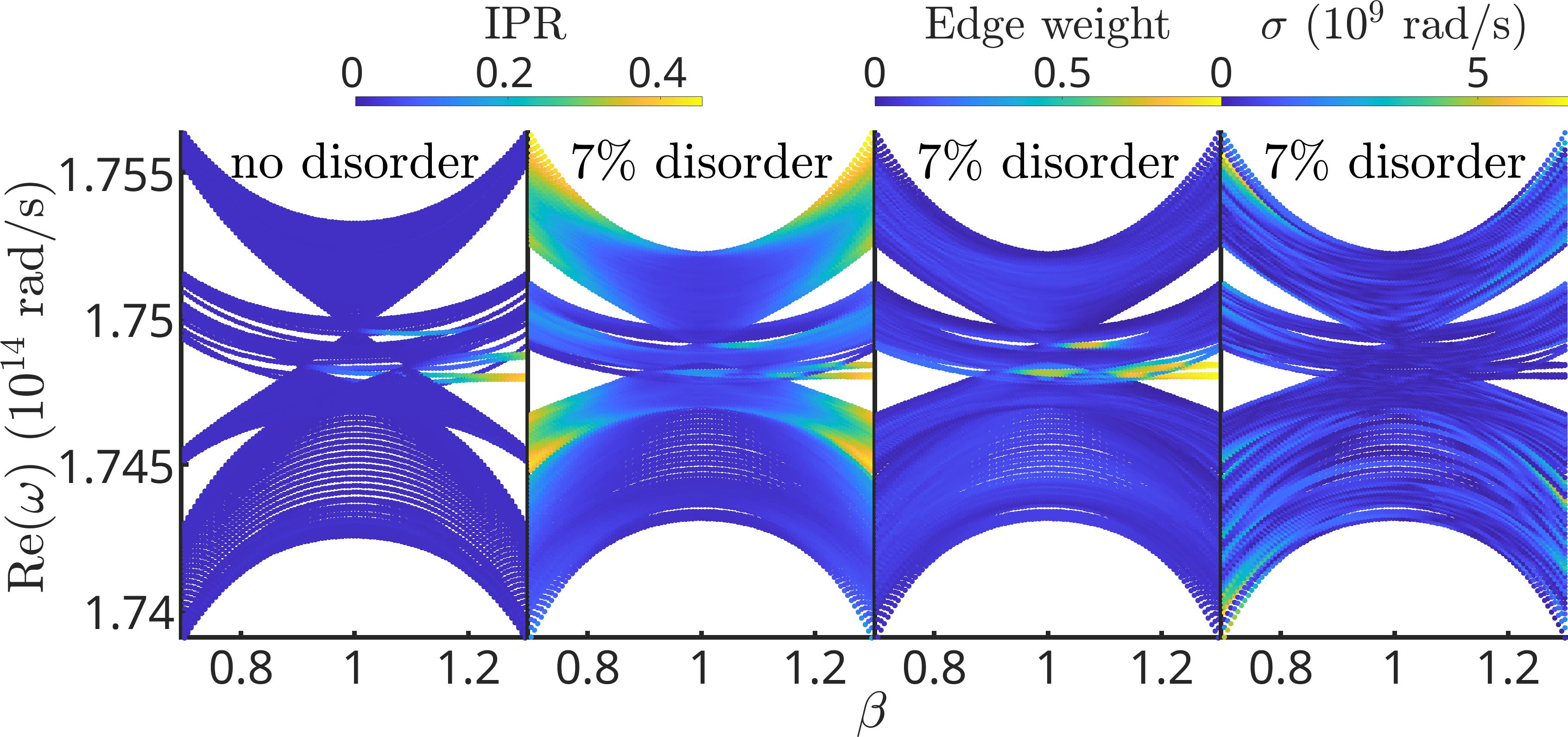}
	\caption{Robustness of the edge modes against disorder. The IP bands of Fig.~~\ref{Fig:dots_beta}(b) are shown for $p = 7 \%$ disorder in terms of IPR values, edge weights, and standard deviations averaged over 100 configurations.}
	\label{Fig:robust_IP}
\end{figure*}

As can be seen in Fig.~\ref{Fig:robust_IP}, also for the IP modes, all edge modes remain identifiable by means of their edge weights, confirming their topological resilience. Notably, the IPR values of previously delocalized bulk modes can increase to a level comparable to, or even exceeding, those of the edge modes. This phenomenon arises because the introduced structural imperfections break the translational symmetry, leading to the formation of Anderson-like localized states and additional defect states within the bandgaps~\cite{Mondragon-Shem2014}. The emergence of Anderson-localized states is particularly prominent within the dipole-inactive bands. Unlike the dipole-active branches, which benefit from long-range interactions mediated by the substrate's SPP modes, the dipole-inactive bands rely on short-range, direct dipole-dipole coupling. This lack of long-range connectivity reduces the spectral resilience of these modes, making them highly susceptible to structural disorder. Consequently, even small spatial perturbations lead to localization as observed by the high IPR values. While this increased spectral density of localized states complicates the visual identification of the topological states, the edge modes, including those for dimerizations $\beta<1$, can still be identified and remain spectrally distinct in the edge weights. Additionally, the edge modes possess a particularly low standard deviation from their ensemble average for the here considered 100 configurations.

\subsection{Examples}\label{sec:RHT}

\subsubsection{Radiative heat transfer}

Here, as an example, we study the impact of the observed effects on the radiative heat transport along an SSH chain~ \cite{Ott2020_2, Wang2023_2, Ott2021, Naeimi2026}. More specifically, we examine the spectrum of the radiative heat flux between the first and last particles of the chain. Based on the established effects induced by the dimerization parameter $\beta$ and the chain-surface distance $z$, we expect the signatures of the observed edge modes to be reflected in the spectral heat flux. The study of the chain-substrate-distance influence on the heat flux accompanied by the study on perturbation complements the results established by Ref.~\cite{Naeimi2026}.

\begin{figure}
	\centering
	\includegraphics[width=0.5\textwidth]{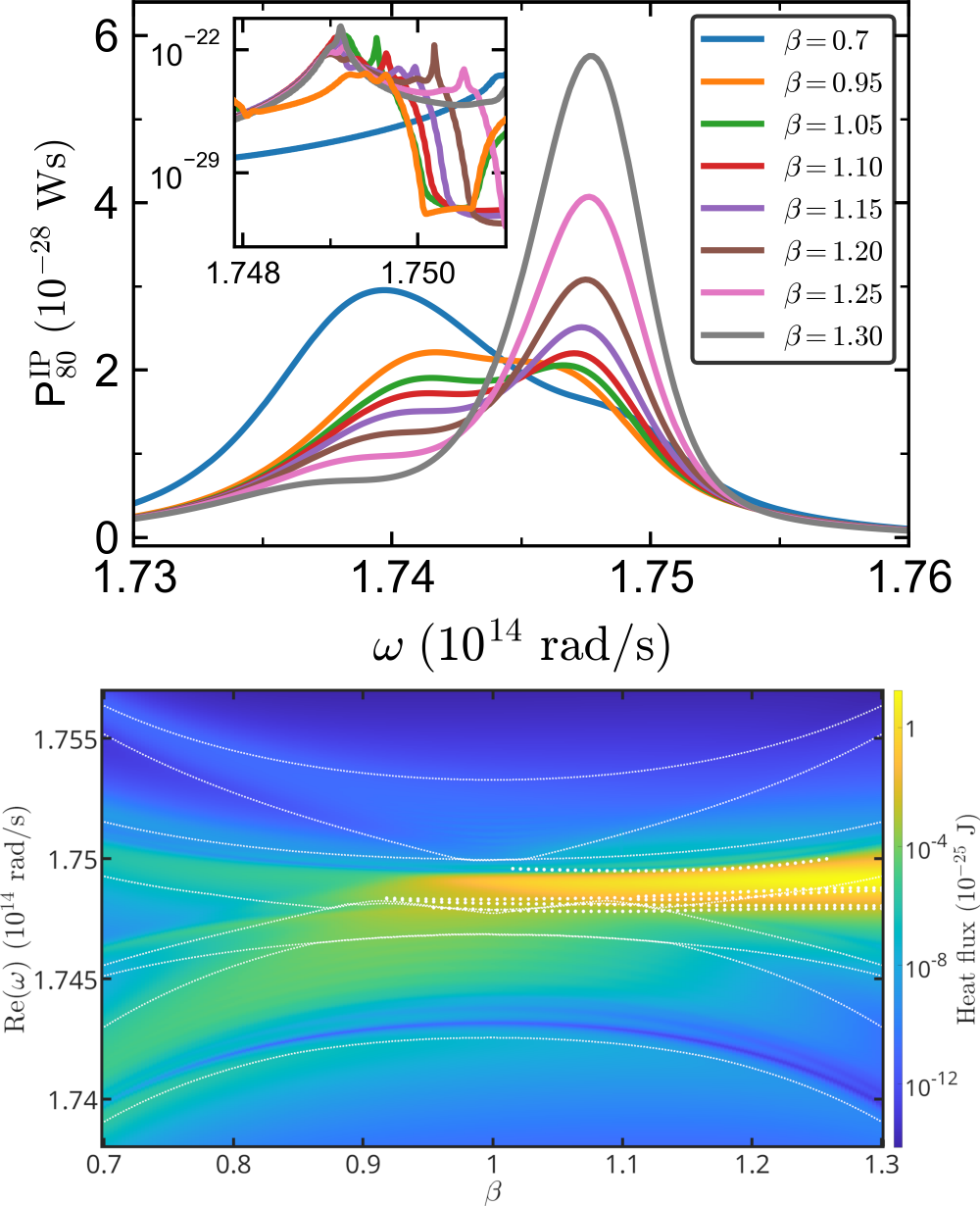}
	\caption{Spectral radiative heat flux between the edges of an 80-particle SSH chain with $n_\text{sub} = 1.36 \times 10^{19}$ cm$^{-3}$ and $z = 500$ nm. Upper panel:  Spectral plot for different $\beta$ values while the inset shows the identical setup with reduced damping constant for the InSb particles by $\Gamma_\text{new} = \Gamma / 100$ to make the edge-mode resonance visible. Lower panel: False-color plot for $\Gamma_\text{new} = \Gamma / 10$ when varying both the frequency and the dimerization. The white lines represent the bands and the dots the edge modes from the eigenmode analysis in Fig.~\ref{Fig:dots_beta}(b).}
	\label{Fig:Spec_beta}
\end{figure}

Starting with the variation of the dimerization $\beta$, we compare our findings for the different edge mode branches in Fig.~\ref{Fig:dots_beta}(b) with our results for the heat flux between the first and last chain particles presented in Fig.~\ref{Fig:Spec_beta} for $n_\text{sub} = 1.36 \times 10^{19}$ cm$^{-3}$ and $z = 500$ nm for the IP direction. The main figure clearly illustrates the peaks generated by edge modes at $\beta>1$, which are particularly pronounced when compared to the reference case of $\beta=0.7$ where no edge modes are observed in Fig.~\ref{Fig:dots_beta}(b). Notably, we also observe a distinct peak at the edge mode frequency predicted for $\beta=0.95$. Because the edge mode frequencies are tightly clustered, we examined whether we can identify the emergence and disappearance of edge mode branches by artificially reducing the damping constant of the InSb particles by a factor of 0.01. This narrowing of the spectral peaks is depicted in the inset of Fig.~\ref{Fig:Spec_beta}. For $\beta > 1.2$, we identify two prominent peaks, corresponding to the two edge mode branches at $\omega = 1.749 \times 10^{14}$ rad$/$s in Fig.~\ref{Fig:dots_beta}(b). We also observe a third peak, which is less intense and appears at higher frequencies initially, but whose frequency shifts toward lower values as the dimerization decreases. Due to the strong spectral decay after the appearance of this peak, we assign it to the upper transversal band. Additionally, a smaller peak appears at slightly lower frequencies, the intensity of which increases as $\beta$ decreases. At $\beta=1.05$, this peak becomes the dominant feature at the frequency of the predicted upper edge mode branch. The lower panel of Fig.~\ref{Fig:Spec_beta} showing the spectral heat flux in the $\omega$-$\beta$ plane allows for a one to one comparison with the eigenmodes from Fig.~\ref{Fig:dots_beta}(b). Note that as pointed out in Ref.~\cite{Naeimi2026}, the observed intensity not only reflects the localization of the edge modes but also their ability to couple to the substrate's SPPs. Therefore, a stronger localization does not necessarily result in a higher peak value unless the coupling between those two modes is supported so that peaks in the heat flux cannot be understood as a direct proof of the existance of edge modes, in general, but rather as supporting evidence. These findings complement the behaviors observed in Fig.~\ref{Fig:dots_beta}(b), further highlighting the impact of the band structure on measurable physical quantities like the heat flux.

\begin{figure}
	\centering
	\includegraphics[width=0.5\textwidth]{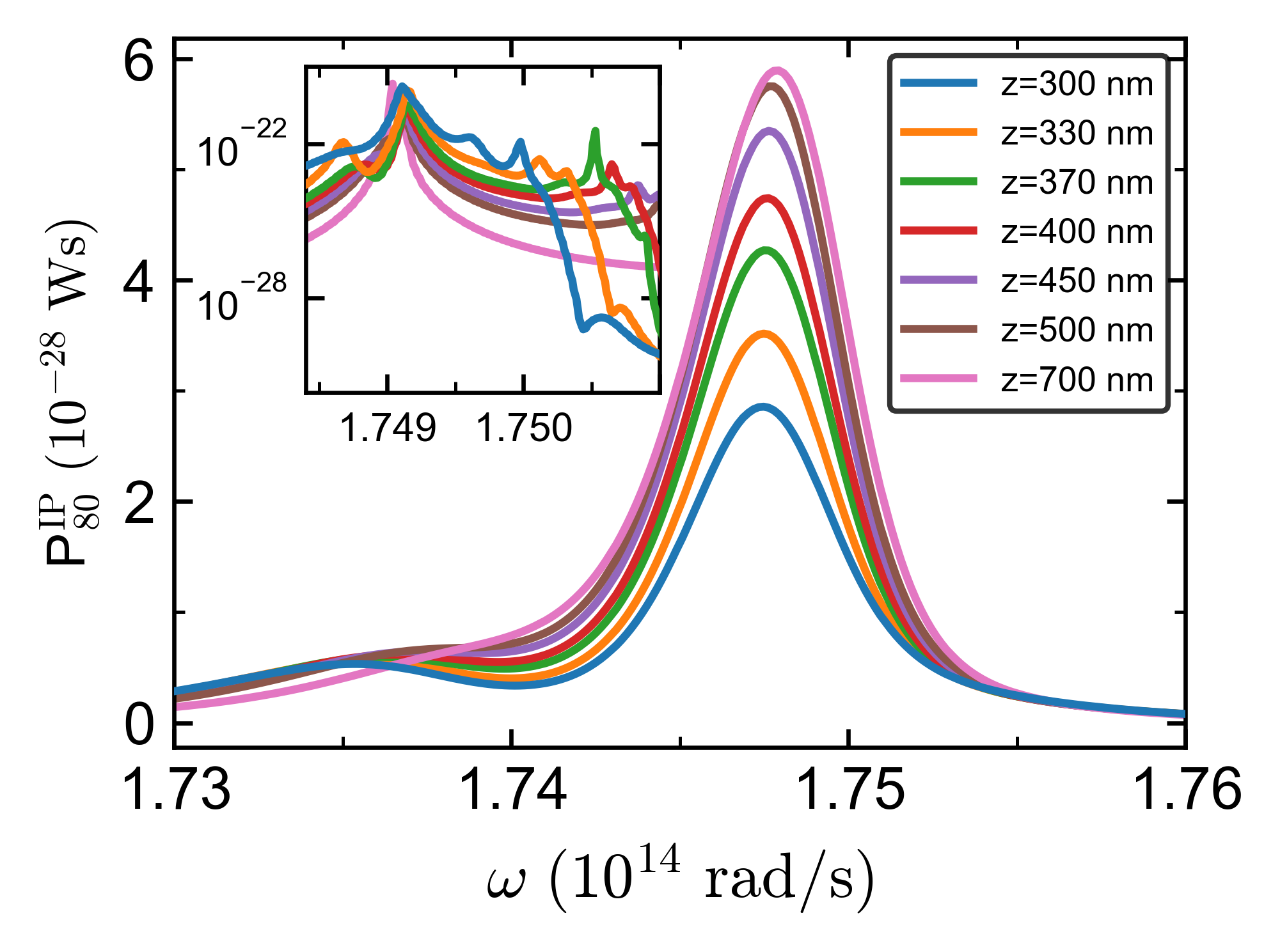}
	\caption{Spectral radiative heat flux between the edges of an 80-particle SSH chain with $n_\text{sub} = 1.36 \times 10^{19}$ cm$^{-3}$ and $\beta = 1.3$ for different $z$ values. Inset: Identical setup with reduced damping constant for the InSb particles by $\Gamma_\text{new} = \Gamma / 100$ to make the edge-mode resonance visible.}
	\label{Fig:Spec_z}
\end{figure}

Following a similar analysis, we compare the spectral heat flux between the first and the last chain particles as illustrated in Fig.~\ref{Fig:Spec_z} for various gap sizes $z$ between the chain and the substrate, keeping $n_\text{sub} = 1.36 \times 10^{19}$ cm$^{-3}$ and $\beta = 1.3$ constant. The main figure in Fig.~\ref{Fig:Spec_z} exhibits a prominent peak around the edge mode frequencies visible in Fig.~\ref{Fig:dots_z}(b), though these modes cannot be fully separated. Note that the maximum peak does not appear at the shortest chain-surface distance. This is again a consequence of the coupling between edge modes and the substrate's SPPs, as pointed out in Ref.~\cite{Naeimi2026} through the dispersion relation in Eq.~\eqref{Eq:DispSPP}, which relates the $k_\perp$ value at the edge mode's frequency to the gap size via the condition $2 k_\perp z \approx 1$. To distinguish between the different edge mode branches, we repeat the process of reducing the InSb damping constant, which is displayed in the inset of Fig.~\ref{Fig:Spec_z}. The two closely separated edge modes predicted for $z < 500$ nm at $\omega = 1.749 \times 10^{14}$ rad$/$s are clearly resolved as two peaks, where the peak at lower frequencies shifts toward lower values as $z$ decreases. This behavior coincides with the redshift of the lower edge mode branch observed in Fig.~\ref{Fig:dots_z}(b). The appearance of the new upper edge mode branch for $z<370$ nm, however, is difficult to distinguish from the vanishing upper edge mode branch seen for $z>370$ nm, as they reside at nearly identical frequencies due to the frequency redshift. Nevertheless, we notice a small shift toward higher frequencies for the peak at $\omega = 1.749 \times 10^{14}$ rad$/$s in the range $300$ nm $<z<370$ nm, which follows the increase of the upper edge mode branch's frequency. For $z>370$ nm, this trend reverses, causing the peak to shift back to lower frequencies and remain constant thereafter. We identify this pattern with the emergence and disappearance of the upper edge mode branches as shown in Fig.~\ref{Fig:dots_z}(b).

\begin{figure}
	\centering
	\includegraphics[width=0.5\textwidth]{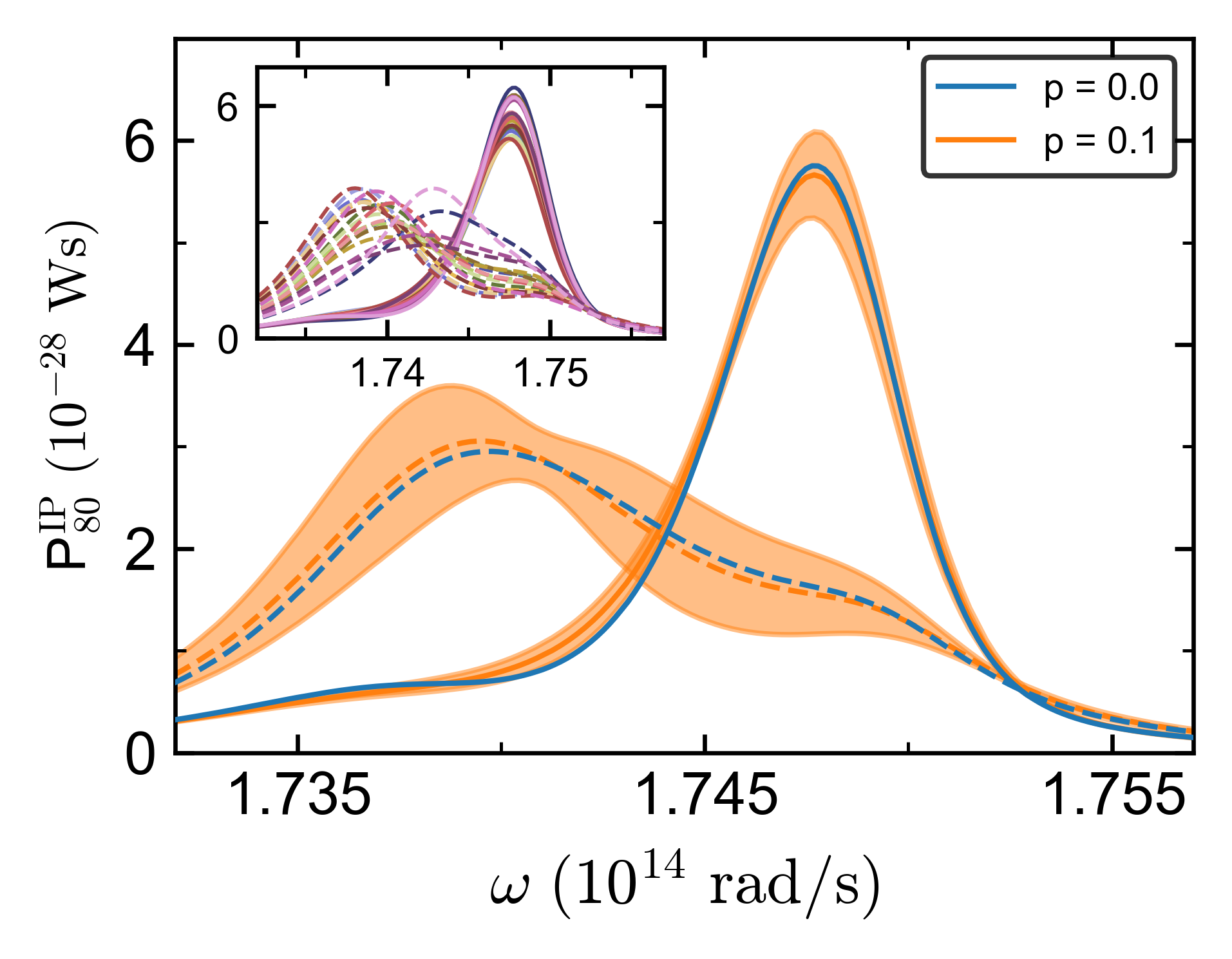}
	\caption{Spectral radiative heat flux between the first and last particle of an 80-particle SSH chain in the vicinity of a substrate for $n_\text{sub} = 1.36 \times 10^{19}$ cm$^{-3}$, $z = 500$ nm, and $\beta = 0.7$ (dashed) as well as $\beta = 1.3$ (solid) with (blue) and without (red) $p=0.1$ disorder. The orange shaded area displays the standard deviation from the ensemble average for 20 configurations. Inset: Spectral radiative heat flux for all 20 configurations.}
	\label{Fig:Spec_robust}
\end{figure}

Finally, we analyze the robustness observed in Fig.~\ref{Fig:robust_IP} for the IP edge modes with respect to the heat flux along the 80-particle SSH chain. The spectral heat flux between the chain ends is displayed in Fig.~\ref{Fig:Spec_robust} for $n_\text{sub} = 1.36 \times 10^{19}$ cm$^{-3}$ and $z = 500$ nm under disorders of $p = 0$ and $p = 0.1$. Note that for $\beta = 0.7$ and $\beta = 1.3$, we fix the chain ends while allowing displacements of the intermediate particles due to disorder. For the disordered case, we computed the spectra over 20 configurations. In Fig.~\ref{Fig:robust_IP} we present the averaged spectra together with the standard deviation. The inset shows the spectra of all individual configurations.

When comparing the results for $\beta = 0.7$ and $\beta = 1.3$, we find that the averaged spectra remain very close to the unperturbed spectrum due to the small disorder amplitude. However the standard deviation is substantially larger across all frequencies for the trivial case $\beta = 0.7$ than for the non-trivial case $\beta = 1.3$. From the individual configurations in the inset of Fig.~\ref{Fig:robust_IP}, it is apparent that in the trivial case the dominant frequency of the heat transfer via band modes is strongly influenced even by a small disorder of $10\%$, thereby causing the large overall standard deviation. In contrast, for the non-trivial case the dominant frequency is relatively robust against disorder due to the resilience of the edge modes. Interestingly, the magnitude of the spectral heat flux can change noticeably across individual configurations, resulting in a pronounced standard deviation close to the edge mode resonance. Thus, the characteristics of the edge modes clearly leave their imprint on the radiative heat flux.

\subsubsection{Local Density of States}

As a second example we consider the electric LDOS at an observation point $\mathbf{r}$. To be precise, we determine solely the contribution of the nanoparticles to the LDOS by subtracting the substrate contribution (as in Ref.~\cite{Naeimi2026}), thus considering the quantity~\cite{Ott2021}
\begin{equation}
	D^{\rm E}(\omega,\mathbf{r}) = \frac{k_0^3}{\pi c}\sum_{i,j = 1}^N{\rm ImTr}\Big\{\alpha\mathds{G}_{0i}^{\rm tot} \boldsymbol{T}_{ij}^{-1} \mathds{G}_{j0}^{\rm tot}\Big\},
	\label{Eq:eLDOS}
\end{equation}
where the $\mathds{G}_{j0}^{\rm tot} = \mathds{G}^{\rm tot}(\mathbf{r}_j,\mathbf{r})$ and
\begin{equation}
	\boldsymbol{T}_{ij} = \delta_{ij}\mathds{1}-(1-\delta_{ij})k_0^2\alphamat\mathds{G}_{ij}^{\rm tot}.
\end{equation}

\begin{figure}
	\centering
	\includegraphics[width=0.5\textwidth]{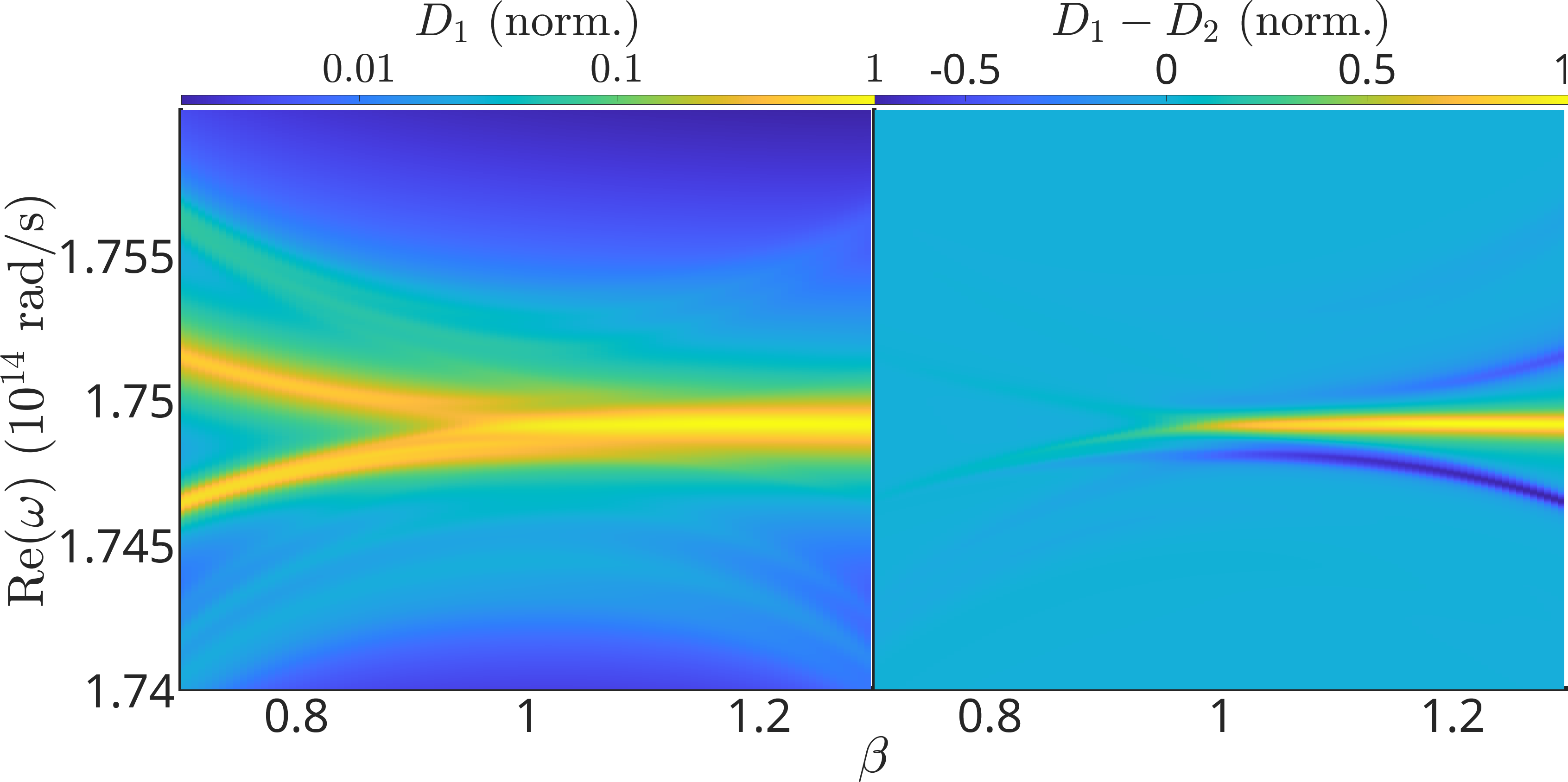}
	\caption{LDOS for different frequencies and dimerizations at $z=200$ nm and with $n_\text{sub} = 1.36 \times 10^{19}$ cm$^{-3}$. The LDOS below the first particle (left panel) normalized to its maximum value and the difference between the LDOS below the first and second particle (right panel) are shown.}
	\label{Fig:LDOS}
\end{figure}

In Fig.~\ref{Fig:LDOS} we show the electric LDOS defined in Eq.~\ref{Eq:eLDOS} for a configuration of $8$ nanoparticles at a distance of 500nm above an InSb substrate with  $n_\text{sub} = 1.36 \times 10^{19}$ cm$^{-3}$ as a function of frequency $\omega$ and dimerization $\beta$. We evaluate the LDOS at positions directly below the first particle and below the second particle of the chain at a distance of $z = 200$nm, i.e., directly midway between the nanoparticle edges and the substrate. The LDOS value below the first particle $D_1$ is shown in the left panel. An enhanced LDOS for the two bands forming the band gap can be observed for $\beta < 0.9$, whereas for $\beta > 0.9$, the edge mode starts to dominate. In the right panel of Fig.~\ref{Fig:LDOS} we show $D_1 - D_2$, i.e., the difference in LDOS below the first particle $D_1$ and below the second particle $D_2$. This difference even more clearly demonstrates how the edge modes dominate the LDOS below the first particle for $\beta > 0.9$, providing further evidence for the existence of edge modes for $\beta < 1$. It should, however, be noted that this is only clearly visible for reduced damping, where the linewidth of the modes becomes small enough to resolve their individual impact. Therefore, applications aiming to probide clear evidence of an edge mode enhanced LDOS for $\beta < 1$ require materials with low damping.

%
%
\section{Conclusion}\label{sec:conclusion}

The band structure of a plasmonic topological SSH chain can be significantly altered when placed in close proximity to a substrate. This interaction enables two interesting mechanisms that introduce additional Zak phases and, consequently, facilitate the emergence of new edge modes or the vanishing of existing ones. These two mechanisms correspond to the hybridization of the two dipole-active bands due to the chain-substrate coupling, as well as the band touching of the two transversal bands. We find that the hybridization is a long-range effect, whereas the band touching is a short range effect. The latter leads to a new band opening which, in analogy to the well-known case at $\beta = 1$, allows for the emergence of new edge modes. Beyond varying the dimerization $\beta$, adjustments to the substrate's doping and the gap size are also capable of significantly altering these band structures and triggering these mechanisms. Additionally, the edge modes remain robust against disorder introduced by slightly displacing the particles along the chain. We also identified signatures of these edge modes and their triggering mechanisms in the spectral heat flux between the chain ends and the LDOS in the vicinity of the first nanoparticle.

This establishes a new approach to engineering topological band structures, not only by tuning the primary parameters of the particle chain, but more importantly, by changing its environment. Our findings facilitate not only the manipulation of inherent properties like the topological phase transition, but also the tuning of the edge mode frequencies. Such control is particularly valuable in scenarios where precise command over edge mode coupling is necessary. Since these effects remain robust against disorder, fabrication tolerances for such geometries are significantly relaxed, as the topological protection remains largely unaffected by structural imperfections.

%
%
\section*{Acknowledgements}

The authors gratefully acknowledge financial support from the Niedersächsische Ministerium für Kultur und Wissenschaft (`DyNano'). Florian Herz acknowledges financial support from the Deutsche Forschungsgemeinschaft under project number 570757245. Simulations were conducted on the the HPC cluster ROSA funded by the Deutsche Forschungsgemeinschaft under INST 184/225-1 FUGG.

\end{document}